\documentclass[conference]{IEEEtran}

\usepackage{textcomp}
\usepackage{pgfplots}
\usepackage{booktabs}
\usepackage{multirow}
\usepackage{dcolumn}
\newlength{\thinline}
\newlength{\thickline}
\usepackage[cmex10]{amsmath}
\usepackage{amssymb}

\usepackage{subfigure}

\usepackage[utf8]{inputenc}
\usepackage[english]{babel}
\usepackage{amsthm}

\usepackage[algo2e]{algorithm2e}
\usepackage{algorithm}
\usepackage{array}

\usepackage{dblfloatfix}
\usepackage[hyphens]{url}
\usepackage{hyperref}
\usepackage{xcolor}
\usepackage{balance}
\usepackage{caption}

 \newcommand{\TMMPP}{\texttt{TMMPP}\xspace}
\newcommand{\hwl}[1]{\textcolor{black}{#1}}
\newcommand{\lgp}[1]{\textcolor{black}{#1}}
\newcommand{\sls}[1]{\textcolor{black}{#1}}

\begin{document}

\title{SoK: Training Machine Learning Models over Multiple Sources with Privacy Preservation}

 \author{
Lushan Song, Guopeng Lin, Jiaxuan Wang, Haoqi Wu, Wenqiang Ruan, Weili Han \\
Laboratory for Data Analytics and Security, Fudan University 
}


\maketitle

\begin{abstract}
Nowadays, gathering high-quality training data from multiple data sources with privacy preservation is a crucial challenge to training high-performance machine learning models. The potential solutions could break the barriers among isolated data corpus, and consequently enlarge the range of data available for processing. To this end, both academic researchers and industrial vendors are recently strongly motivated to propose two main-stream folders of solutions mainly based on software constructions:
1) Secure Multi-party Learning (MPL for short); and 2) Federated Learning (FL for short). 
The above two technical folders have their advantages and limitations when we evaluate them according to the following five criteria: security, efficiency, data distribution, the accuracy of trained models, and application scenarios. 

Motivated to demonstrate the research progress and discuss the insights on the future directions, we thoroughly investigate these protocols and frameworks of both MPL and FL.
At first, we define the problem of \underline{T}raining  machine learning \underline{M}odels over \underline{M}ultiple data sources with \underline{P}rivacy \underline{P}reservation (\TMMPP for short). Then, we compare the recent studies of \TMMPP from the aspects of the technical routes, the number of parties supported, data partitioning, threat model, and machine learning models supported, to show their advantages and limitations. Next, we investigate and evaluate five popular FL platforms. Finally, we discuss the potential directions to resolve the problem of \TMMPP in the future.
\end{abstract}

\section{Introduction}
\label{introduction}
In the era of big data, almost all online activities are driven by data. The data, which would bring us enormous benefits, become the key to current business competition. In addition, data are pushing the advances of machine learning, which have brought us breakthroughs in various scenarios, such as medical diagnosis~\cite{esteva2017dermatologist,fakoor2013using}, image classification~\cite{rastegari2016xnor}, and natural language processing~\cite{gu2021domain}. In the aforementioned widely-applied scenarios, massive amounts of high-quality data are usually a dominating factor affecting the performance of machine learning models. 

However, researchers face a tough predicament where the needed data are hard to collect, merge or share directly. That is because the needed data are usually decentralized among multiple data owners and might contain much sensitive or private information, which is protected by recently released privacy protection policies or regulations~\cite{ruan2021snp}, such as General Data Protection Regulation (GDPR for short)~\cite{voigt2017eu} of European Union.

Therefore, how to utilize the decentralized data in multiple data owners to efficiently train high-performance machine learning models  with privacy preservation is now a crucial challenge. 
\hwl{In this challenging scenario, when vendors want to train the models, the data owners hold valuable data under their control. The potential risk of sensitive information leakage hereby incurs regulations enforced by related laws, thus further leading to the phenomenon of isolated data sources.}



In this paper, we try to deeply investigate these current advances of resolving the problem of \underline{T}raining  machine learning \underline{M}odels over \underline{M}ultiple data sources with \underline{P}rivacy \underline{P}reservation, referred to as \TMMPP,
from two technical folders: 1) Secure Multi-party Computation (MPC for short) based machine learning~\cite{bogdanov2008sharemind,demmler2015aby,mohassel2017secureml,ruan2022private}, whose architecture is typically the Peer-to-Peer model. We named it Secure Multi-party Learning (MPL for short) in this paper. And 2) Federated Learning (FL for short)~\cite{mcmahan2016federated}\cite{konevcny2016federated1}\cite{konevcny2016federated2}\cite{mcmahan2017communication}, whose architecture typically follows the Client-Server model. Both these two folders of solutions usually depend on software constructions. Thus, many researchers will concern about which one is a mainstream way to resolve \TMMPP, or which one should be chosen
by users to resolve a concrete scenario of \TMMPP. 
Note that researchers and engineers also propose another technical solution based on the trusted execution environment~\cite{ohrimenko2016oblivious}, which requires the support of additional hardware and we do not investigate in this paper. In this solution, researchers use the technologies of a trusted execution environment to train the models over a central data source from distributed locations. The privacy is preserved by the trustworthiness of the data process environment, where executors usually can 
only obtain the final results and cannot know the details of raw data.

MPC, firstly proposed by Andrew C. Yao in 1982 for the millionaire problem~\cite{yao1982protocols}, is extended into a general definition for any polynomial computable function in 1986~\cite{yao1986generate}.
The identification of MPC is groundbreaking, creating opportunities in which semi-trusted parties can jointly calculate results of mutual interest without a trusted third party.
Two millionaires can use Yao's protocol to figure out who is richer without revealing their concrete assets to each other or a third party.
MPC provides a security mechanism by which a group of mutually distrusting parties can jointly compute functions without revealing any private information beyond the function result. Especially, MPC can be implemented without a trusted third party. 
With the support of MPC protocols, researchers have developed numerous privacy-preserving machine learning frameworks to keep the training data private in recent years.
In real-world scenarios, however, the expensive communication overhead and computation complexity of their underlying protocols are still two severe issues before these frameworks become practical.

In 2016, Google presented the concept of  FL~\cite{konevcny2016federated1}\cite{konevcny2016federated2}, which made distributed privacy-preserving machine learning be a hot topic once again.
FL enables data owners to efficiently train machine learning models without collecting raw data.
In the FL framework proposed by Google, each data owner trains a model locally with their local data, and uploads the intermediate training results such as model updates rather than raw data to a centralized server.
Briefly, FL iteratively executes the three steps as follows: 1) the centralized server sends the current global model to the parties or a subset of them;
2) each party trains a local model to tune the global model accepted from the centralized server using its local data; 
3) the centralized server updates a new global model via aggregating the local model updates from parties.
That is, the data owners and the centralized server work together by exchanging local model updates and global parameters. Thus, FL usually has lower communication overhead and computation complexity than those of MPL.
In the past five years, FL became a hot technology to resolve the problem of \TMMPP. Lots of practical frameworks~\cite{smith2017federated}\cite{cheng2019secureboost}\cite{liang2020think}\cite{xu2019verifynet} and applications~\cite{wang2019edge}\cite{kim2019blockchained}\cite{nishio2019client}\cite{zhao2020local} extended 
from the FL framework by Google in 2016 have been presented.

\subsection{Related Surveys}
The problem of \TMMPP has been studied for decades. Academic researchers are motivated to present protocols or frameworks, while industrial vendors build up practical platforms. 
Several related surveys have also been conducted concerning the topic of MPL and FL, respectively.

Archer et al.~\cite{archer2016maturity} investigated the state-of-the-art secure computation technologies, described three paradigms of programmable secure computation, including Homomorphic Encryption (HE for short), Garbled Circuit (GC for short), and Secret Sharing (SS for short).
Then they evaluated schemes based on existing applications and benchmarks. Mood et al.~\cite{mood2016frigate} presented a brief survey about existing secure computation compilers, and analyzed these compilers' correctness. Hastings et al.~\cite{hastings2019sok} surveyed general-purpose compilers for MPC, which focused on their usability features, along with container-based virtual environments to run example programs.

Yang et al.~\cite{yang2019federated} provided a seminal survey of existing works on FL, generally introduced the definition, architecture and techniques of FL, and further discussed its potential in various applications. Li et al.~\cite{li2019federateda} discussed the challenges faced by FL and presented a broad overview of current works about FL around these challenges. Finally, they outlined several open directions as the future work.
A comprehensive survey written by Kairouz et al.~\cite{kairouz2019advances}, summarized the recent advances and challenges of FL from various research topics.
The literature~\cite{li2019federatedb} The literature surveyed FL systems while providing a categorization according to six different aspects and presenting the comparison among existing FL systems.
Lim et al.'s~\cite{lim2020federated} survey highlighted the issues and solutions regarding the FL implementation to Mobile Edge Computing.

When we consider the rapid development of the solutions to \TMMPP, however, a comprehensive and systematic survey, which covers the main-stream technical folders, i.e., MPL and FL, based on cryptography technologies, is still absent so far.
Thus, we are motivated to conduct such a literature review of the technical routes, frameworks, and platforms of \TMMPP. 
Our study would help researchers choose suitable \TMMPP frameworks and platforms for various scenarios, further identify research gaps, and improve the weaknesses of the approaches.
\subsection{Our Contributions}
In this paper, our main contributions are as follows:

\begin{itemize}
     \item We present a definition of the problem of \TMMPP, and list the challenges faced by \TMMPP. The challenges include the security challenge, efficiency challenge, and statistical challenge, where the statistical challenge is mainly faced by FL. 
    \item We investigate the state-of-the-art studies proposed to resolve \TMMPP, and classify the various frameworks into seven technical routes based on their underlying techniques. 
    Then, we provide a systematic analysis of the characteristics among seven technical routes, along with the differences between different MPL frameworks and FL frameworks, respectively. 
    We point out that MPL frameworks generally are more secure than FL frameworks and are not sensitive to how the data are distributed among the parties while FL frameworks usually have lower communication overhead. In addition, we argue that the SS-based MPL frameworks, Mixed-protocol based MPL frameworks, and DP-based FL frameworks, SC-based FL frameworks will respectively become the mainstream of the solutions of MPL and FL.
      \item Because MPL platforms generally involve one technical route whereas FL platforms are basically enterprise-level with more complex designs and suitable for a variety of scenarios, we sketch five popular FL open-source platforms (i.e. \texttt{FATE}, \texttt{TFF}, \texttt{PaddleFL}, \texttt{Pysyft}, \texttt{coMind}). We compare them using a series of indicators based on the information provided by the official websites and open-sourced codes, including data partitioning, privacy technologies, ways of deployment, hardware, and visual interface. 
      Furthermore, we evaluate the efficiency and accuracy of these five open-source platforms over logistic regression (LOR for short) and neural networks (NN for short).
    
    \item We present a comprehensive overview of the history to solve \TMMPP and discuss its future directions, including improving security, improving efficiency, supporting more data distributions and others, such as improving dynamic scalability, supporting more complex models, and tolerating system heterogeneity.

\end{itemize}


\section{Problem Definition and Challenges}
\label{sec:pc}
In this section, we define the problem of \TMMPP. Then, we introduce the security model that we considered. Finally, we identify the major technical challenges when solving the problem of \TMMPP.

\subsection{Problem Definition}
\label{problem}

\begin{figure}[h]
\centering
\includegraphics[scale=0.7]{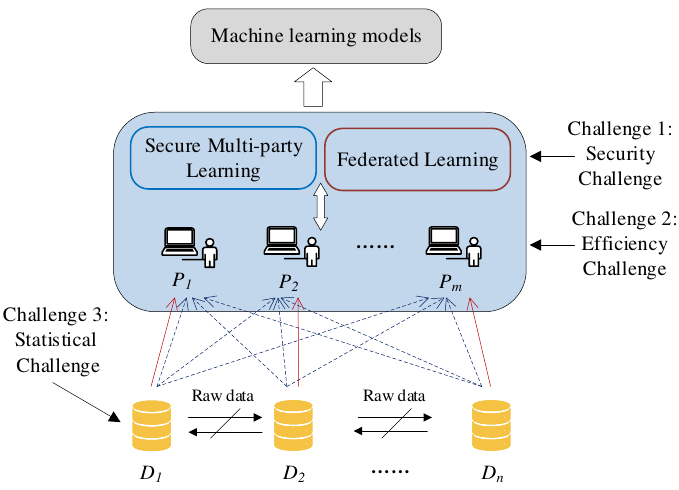}
\caption{Scenarios of training machine learning models over multiple data sources with privacy preservation. $\mathcal{D}_i$ denotes the raw data. $\mathcal{P}_i$ denotes the party. The red solid lines represent that parties are data owners who hold the raw data, and the blue dotted lines represent that data owners outsource computations to parties.}
\label{fig.problem}
\end{figure}

As shown in Figure~\ref{fig.problem}, when we consider the following scenario: a group of mutually distrusting data owners plans to jointly train a machine learning model over their owned data with security assurance. The distribution of data among the data owners can be horizontally partitioned (i.e., data are distributed by samples) or vertically partitioned (i.e., data are distributed by features). These data owners communicate with or without a centralized server during the training but do not disclose their raw data.
It is noteworthy that the process of machine learning usually consists of two phases: the training phase, and the inference phase. But we here only focus on the secure training phase, which is more generic since the secure inference is naturally implied when training is done.

Before we present a formal definition of the above problem, we define the following notations. There are two roles: the data owner who holds the raw data and the party who actually participates in the training process to perform computations. 
In MPL, the data owner may outsource computations of the training to some other parties, while, in FL, the data owner generally participates in the training, that is, the data owner has the same role as the party.
Without loss of generality, let $\lbrace\mathcal{D}_1, \mathcal{D}_2, ... , \mathcal{D}_n \rbrace$ respectively be the raw datasets, which are held by $n$ data owners, $\lbrace\mathcal{P}_1, \mathcal{P}_2, ... , \mathcal{P}_m \rbrace$ be $m$ parties. We use $\mathcal{Mo}$ to denote a machine learning model cooperatively trained by these $m$ parties.

Then, the problem of \TMMPP can be stated as follows:

\textbf{Input:} Each data owner takes its owned raw data $\mathcal{D}_i$ as input. 

\textbf{Output:} The output is a global model $\mathcal{M}$, jointly trained by all  $\mathcal{P}_i$ without exposing any information about the raw data to others in the training process.

\subsection{Security Model}
In general, we can define the security model from two aspects: the behavior of the corrupted parties and the number of corrupted parties.

According to the aspect of the behavior of the corrupted parties, there are two kinds of security models, i.e., semi-honest and malicious corruption. The difference between semi-honest and malicious corruption is whether corrupted parties follow the protocol or not.
Semi-honest (also known as honest-but-curious or passive) corrupted parties in the solutions to \TMMPP attempt to gain as much information as possible from the data they get from intermediate results, but they do not deviate from the protocol specification. Malicious (also known as active) corrupted parties can break the protocol arbitrarily, such as by sending incorrect messages to other parties. 

In addition, according to the number of parties who are corrupted in some way, there are two kinds of security models, i.e., honest/dishonest majority.
The difference between an honest majority and a dishonest majority is whether the number of corrupted parties is less than half of the number of total parties or not.
For the two-party setting, we do not consider an honest majority or a dishonest majority security model.

Combining security assumptions from the above two aspects, there are four combinations. The strongest security assumption is an honest majority with semi-honest corruption and the weakest security assumption is a dishonest majority with malicious corruption.


\subsection{Challenges}
\label{sec.challenge}
We summarize three key challenges: security challenges, efficiency challenges, and statistical challenges, which we will meet when we solve the problem of \TMMPP.


\subsubsection{Security Challenges}\lgp{Each} party and the centralized server may not be fully trusted (following the assumption of the framework's security model). I.e., some corrupted parties can infer certain private information from the information obtained, even somehow impose attacks on private information or interfere with the normal execution of the training algorithms. 

Relying on cryptographic primitives such as HE, GC, and SS, each solution of MPL ensures that under the assumption of the security model, no other sensitive information will be revealed beyond the computation result. However, FL aims to learn the global machine learning model under the constraint that each party's data are stored and processed locally, with only \lgp{model updates} being \lgp{uploaded to} a centralized server. However, the \lgp{model updates} also contain \lgp{some} sensitive information, \lgp{which} may be leaked~\cite{melis2019exploiting}. That is, the privacy of parties is under threat if there is a powerful attack on the \lgp{model updates} during the training of machine learning models. 
Furthermore, MPL can potentially keep the model private to the parties because the final global model held by each party can be in the form of ciphertext. In contrast,  parties in FL will definitely get specific information about the model (even if it is may not the final global model) because some parties can obtain the updated global model from the centralized server in each iteration of FL.

\smallskip
\noindent\textbf{Insight:}
\emph{The solutions of MPL generally more secure than \lgp{the} solutions of FL.}
\smallskip

\subsubsection{Efficiency Challenges} In the solutions of MPL, the costs of communication and computation greatly depend on their underlying protocols. For instance, HE-based protocols usually lead to high computation complexity, while GC-based protocols generally lead to expensive communication overhead. To improve the efficiency of the solutions of MPL, a trade-off between communication overhead and computation complexity is critical.
Besides, training machine learning models over multiple data sources usually involves several, even a considerable number of data parties. E.g., in a typical scenario of FL~\cite{mcmahan2017communication}, there are thousands of parties. For the communication among parties that can be coordinated by a centralized server in FL, the communication overhead is much smaller than the Peer-to-Peer form of MPL, especially when the number of parties is very large. At the same time, in FL, each party utilizes its own plaintext data to train a local model, but in MPL, training is performed on ciphertext, which leads to higher computational complexity.
Due to the communication among parties can be coordinated by a centralized server in FL, the communication overhead of FL is much smaller than the Peer-to-Peer form of MPL, especially when the number of parties is vast and in the IID (independent and identically distributed)) setting. At the same time, in FL, each party utilizes its own plaintext data to train a local model. However, in MPL, training is performed on ciphertext, leading to higher computational complexity.

\smallskip
\noindent\textbf{Insight:} \emph{FL is usually more efficient than MPL when the number of parties is enormous in the IID setting.}
\smallskip

\subsubsection{Statistical Challenges}
The data held by data owners are usually generated or collected in a non-IID (\lgp{non-}independent and identically distributed) manner, i.e., the data may not be independent or have a distinct distribution, even neither independent nor identically distributed. Furthermore, statistical challenges are usually only considered in a horizontally partitioned setting.

In the solutions of MPL, suppose data owners (or parties) use SS protocols to share their owned data with other data owners (or parties) before training, then statistical challenges could have negligible effects. Because after sharing their data, the distribution and size of the data held by each party are the same.
Nevertheless, in \lgp{the} solution of FL, each party trains a local model utilizing their own data. So the non-IID data usually make it challenging to train a high-accuracy machine learning model. 
For instance,  the model's accuracy will drop drastically in the non-IID setting compared with the IID setting. 
Therefore, the non-IID data do not actually impact the training of MPL. However, FL needs to customize different methods for different data distribution.

\smallskip
\noindent\textbf{Insight:}
\emph{MPL is not sensitive to the distribution of the data held by the parties. }



\section{Frameworks}
\label{sec:frameworks}
In this section, we summarize the state-of-the-art academic works to resolve \TMMPP, sketch the different technical routes, and analyze the advantages and disadvantages of these technical routes.

Taking the fundamental privacy technologies utilized during the machine learning training process into consideration, we divide the MPL frameworks into four categories, including HE-based MPL frameworks, GC-based MPL frameworks, SS-based MPL frameworks, and Mixed-protocol based MPL frameworks.
Besides, we divide FL frameworks into three categories according to the type of privacy mechanisms, including Non-cryptographic FL frameworks, DP-based FL frameworks, and SC-based FL frameworks. 
Note that Appendix \ref{sec.ml} shows a brief introduction to machine learning models commonly trained in the MPL and FL frameworks.

\subsection{Background Knowledge of Underlying Primitives and Protocols}
\subsubsection{HE (Homomorphic Encryption)}
HE~\cite{rivest1978data} is a form of encryption that can be used to perform a specific algebraic operation on ciphertext directly, without decrypting it and knowing any information of the private key.
Then it generates an encrypted result, whose decryption result exactly matches the result of the same operation performed on plaintext.
Currently, HE is generally categorized into three types~\cite{acar2018survey}, including Partially Homomorphic Encryption (PHE)~\cite{rivest1978method}\cite{elgamal1985public}, Somewhat Homomorphic Encryption (SWHE)~\cite{boneh2005evaluating} and Fully Homomorphic Encryption (FHE)~\cite{van2010fully}\cite{gentry2009fully}\cite{smart2010fully}\cite{coron2011fully}. 
Here, PHE allows only performing one type of operation (either addition or multiplication) with unlimited times, e.g., Additively Homomorphic Encryption (AHE) of Paillier scheme~\cite{paillier1999public}, which only support performing the addition operation.
In addition, SWHE and FHE can be used to execute both addition and multiplication operations among ciphertexts. 
SWHE can perform certain types of operations within a limited number of times, while FHE can handle all the operations without the restriction of execution times. 
However, the computation of FHE is much more expensive (with higher complexity) than SWHE and PHE. HE executes  a  constant round of interaction in the frameworks.

\subsubsection{OT (Oblivious Transfer)}
OT, as the basis of MPC, can securely transfer values among parties.
Here, we only introduce the 1-out-of-2 OT where the sender $\mathcal{S}$ has two messages $m_0$ and $m_1$, and the receiver $\mathcal{R}$ owns a selection bit $\theta$.
After the execution of OT protocol, the receiver can obtain one of the two messages $m_\theta$ from the sender corresponding to his selection bit, while he learns nothing about $m_{1-\theta}$ and the sender learns no information about $\theta$. One of the implementations of 1-out-of-2 OT in the semi-honest security model is based on the public key. The sender $\mathcal{S}$ generates a public-private key pair ($s_k$, $p_k$) and two random values $x_0$ and $x_1$, then send $p_k$, $x_0$ and $x_1$ to $\mathcal{R}$. $\mathcal{R}$ generates a random value $r$ and calculates $v = x_\theta + Enc_{p_k}(r)$ where $Enc$ refers to the encryption algorithm, and send $v$ to $\mathcal{S}$. After that, $\mathcal{S}$ calculates $r'_i = Dec_{s_k}(v - x_i)$, where $Dec$ refers to the decryption algorithm, $i \in \{0,1\}$, and send $m'_i = r'_i + m_i$ to $\mathcal{R}$. Finally, $\mathcal{R}$ calculates $m_\theta = m'_\theta - r$.

To improve the efficiency, OT extensions are proposed \cite{beaver1996correlated}\cite{ishai2003extending}.
It can combine messages into a long bit string and apply a few basic oblivious transfers to finish massive oblivious transfer tasks.
To cope with messages with specific distributions, Asharov et al.~\cite{asharov2013more} present further efficiency improvement that handles correlated inputs by COT and random inputs by ROT.

\subsubsection{GC (Garbled Circuit)}
GC~\cite{yao1986generate}\cite{huang2011faster}, also known as Yao's garbled circuit, is an underlying method of secure two-party computation, originally proposed by Andrew Yao. GC provides an interactive protocol for two parties (a garbler and an evaluator) obliviously evaluating an arbitrary function which is represented as a Boolean circuit.  


The construction of classical GC includes three main phases: garbling, transferring, and evaluation. Firstly, for each wire $i$ of the circuit, the garbler generates two random strings $k_i^0$ and $k_i^1$ as labels to represent the two possible bit values ``0'' and ``1'', respectively. 
For each gate in the circuit, the garbler creates a truth table.  Each output of the truth table is encrypted using two labels corresponding to its input.
This is done by the garbler to choose a key derivation function that generates symmetric keys using the two labels. Then the garbler permutes the rows of the truth table. After the garbling phase, the garbler transfers the garbled tables, together with the input wire labels corresponding to his input bits, to the evaluator. Moreover, the evaluator acquires the labels corresponding to her input securely by Oblivious Transfer~\cite{beaver1996correlated}\cite{ishai2003extending}\cite{asharov2013more}.
With the garbled tables and labels of the input wires, the evaluator is in charge of decrypting the garbled tables iteratively, until getting the final result of the function.

The total communication overhead of the GC protocol is proportional to the size of the circuit. But regardless of the depth of the circuit and the functionality, GC executes in a constant round of interaction.
Recently, researchers have proposed several variants of the original GC protocol to improve the performance of GC. 
The point-and-permute mechanism proposed by Beaver et al.~\cite{beaver1990round} reduced the number of encryptions. The free XOR protocol~\cite{kolesnikov2008improved} enabled XOR operations performed locally without interaction, which improved the computational efficiency of XOR gates and reduced the size of garbled circuits. Pinkas et al.~\cite{pinkas2009secure} proposed a garbled-row reduction method 4-2 GRR to reduce the size of a garbled table from four to two ciphertexts. Nerveless, it is not compatible with free XOR. Half gates~\cite{zahur2015two} reduced the number of ciphertexts from four to two with free XOR in AND gates.
However, GC-based protocols employ a complex mechanism to transform a function into a Boolean circuit. And these protocols calculate a function bit-by-bit, which leads to significant overhead in communication costs for several operations, especially multiplication.



\subsubsection{SS (Secret Sharing)}
The main idea of SS is to break a secret value into multiple shares, each of which is held by a party.
As each party only owns part of the secret value, multiple parties can execute the operations, e.g., additions and multiplication, without leaking the raw data, then cooperate to reconstruct the result of operations.

Commonly used SS mechanisms include additive SS~\cite{bogdanov2008sharemind} and Shamir's SS~\cite{shamir1979share}. Additive SS refers that the sum of the shares as the secret value. So the secret value can be reconstructed by simply adding all the shares together. In Shamir's SS, the shares are constructed according to a randomized polynomial, and the secret can be reconstructed by solving this polynomial with Lagrange interpolation. As a threshold protocol, Shamir's SS ensures that knowledge of at least the threshold $k$ ($k<n$, $n$ is the number of parties) shares can reconstruct the secret, and nothing can be inferred about the secret if you obtain fewer than $k$ shares. 

Typical SS-based protocols, e.g. GMW (Goldreich-Micali-Wigderson)~\cite{DBLP:conf/stoc/GoldreichMW87}, BGW (Ben-Or-Goldwasser-Wigderson)~\cite{DBLP:conf/stoc/Ben-OrGW88}, RSS (replicated secret sharing)~\cite{benaloh1988generalized}, and SPDZ~\cite{damgaard2012multiparty} are suitable for multiple parties.
    

\noindent\underline{GMW protocol} is a secure multi-party computation protocol, allowing an arbitrary number of parties to securely compute a function that can be represented as a Boolean circuit or Arithmetic circuit. 
All parties share their inputs using the XOR-based SS scheme~\cite{demmler2015aby} and compute the result gate-by-gate.
Since the XOR operation is associative, the XOR gates in the circuit can be evaluated locally by XORing the shares separately without any communication among the parties. And the costs of local calculations of each party can be ignored.
As to AND gates, parties require running an interactive protocol using 1-out-of-4 OT or multiplication triples~\cite{beaver1991efficient} to securely evaluate each gate. The AND gates of the same layer in the circuit can be computed in parallel.
The performance of the GMW protocol depends on the depth of the circuit. 
Unlike GC, GMW based variants do not require generating truth tables, only XOR and AND operations are required for calculation, and allow to pre-compute all symmetric cryptographic operations in the offline phase.
Therefore, GMW usually achieves good performance in low-latency networks~\cite{schneider2013gmw}.



\noindent\underline{BGW protocol} can securely evaluate both Boolean circuit and Arithmetic circuit for more than two parties.
In general, parties initially share their inputs using Shamir's SS, then compute the result gate-by-gate.
For the addition gates in the circuit, the computation can be performed locally without communication among parties, while for the multiplication gates, parties require interactions. 
Rather than using OTs or multiplication triples to calculate results, BGW relies on polynomial multiplications and degree reductions.
Note that BGW requires an honest majority setting.
BGW protocol can be against semi-honest corrupt parties for up to $t < n/2$  and against malicious corrupt parties for up to $t < n/3$, where $t$ is the number of corrupted parties and $n$ is the number of parties.

\noindent\underline{RSS protocol} is essentially a variant of additive SS, which starts with additive SS but sends more than one share to each party. More specifically, take three-party protocol as an example, a secret value $x$ is shared by additive SS into three shares $x_1,x_2,x_3$ such that the sum of them equals $x$. These shares are distributed as pairs $(x_1,x_2),(x_2,x_3),(x_3,x_1)$, where each party holds one of these pairs. As any two parties can reconstruct the secret, this protocol can tolerate up to corruption. 
Araki et al.~\cite{araki2016high} presented a semi-honest three-party RSS protocol. 
In this protocol, each party holds shares pair $(a_1,a_3-x),(a_2,a_1-x),(a_3,a_2-x)$, separately, where the $a_i$ is a random share of zero, i.e. $a_1+a_2+a_3=0$. Zero sharing can be performed using pseudo-random number generation with a set of keys shared between every pair of parties without any communication. Moreover,  the random shares of zero are the only random values needed by the multiplication subprotocol designed by Araki et al.  In the multiplication sub-protocol,  each party sends only one ring element.   Therefore,  this protocol has very low communication: exactly a single round communication per multiplication gate and no communication for addition gates.

\noindent\underline{SPDZ protocol} aims to the dishonest majority multi-party computation with malicious security model, supporting more than two parties. 
SPDZ is an authenticated secret sharing mechanism~\cite{evans2017pragmatic}, split into offline and online phases, where the offline phase is used to generate multiplication triples and perform expensively public-key machinery, and the online phase purely uses cheap, information-theoretically secure primitives.
The main idea of SPDZ is to use an unconditionally secure  message authentication code (MAC for short) to protect secret values from being corrupted by a malicious adversary. More concretely, each party holds pair $(x_i,\gamma(x)_i)$, where $x_i$ is additive share of secret value $x$, $\gamma(x)_i$ is additive share of $\gamma=\alpha(x+\eta)$, and $\eta$ is public value among parties. Moreover, $\gamma$ is the MAC authenticating $x$ under the global key $\alpha$ and $\alpha$ is also distributed in each party as additive share.
SPDZ can defend against malicious corrupted parties for up to $t < n$.

\sls{\subsubsection{DP (Differential Privacy)} Differential Privacy (DP for short)~\cite{dwork2006calibrating}\cite{dwork2008differential}\cite{dwork2014algorithmic}, which aims to add random noise to mask the contribution of any individual user~\cite{chaudhuri2009privacy}\cite{song2013stochastic}~\cite{abadi2016deep}, is a privacy technique with strong information-theoretic guarantees.}

\sls{A randomized algorithm $M: D \to R$ with a domain $D$ and range $R$ satisfies $\epsilon-$differentially private if for any two adjacent datasets $d, d' \in D$ and any subset of outputs $S \subseteq R(M)$, it holds that:
$$Pr(M(d) \in S) \le e^\epsilon Pr(M(d') \in S$$
where $\epsilon$ refers to the privacy budget which limits the bounds of privacy loss.    
Note that two datasets $d$ and $d'$ are adjacent if $d'$ can be formed by adding or removing all the records of a single user from $d$. This notion of DP is referred as user-level DP~\cite{mcmahan2017learning}.
This differs from the adjacency defined by the typical DP where $d$ and $d'$ are adjacent if they differ only in a single record.}

\subsection{Secure Multi-party Learning}
\label{sec.mpc}
Many MPL frameworks have been proposed that utilize one or more cryptographic protocols or primitives, including HE, GC, and SS, to train machine learning models over multiple parties with privacy preservation. 
Since each of these cryptographic protocols owns its characteristics including trade-offs, the frameworks relied on them own their advantages and disadvantages. 

\subsubsection{HE-based MPL Frameworks}
\label{secsec.HE_MPC}



\begin{figure}[ht]
    \centering
    \includegraphics[scale=0.7]{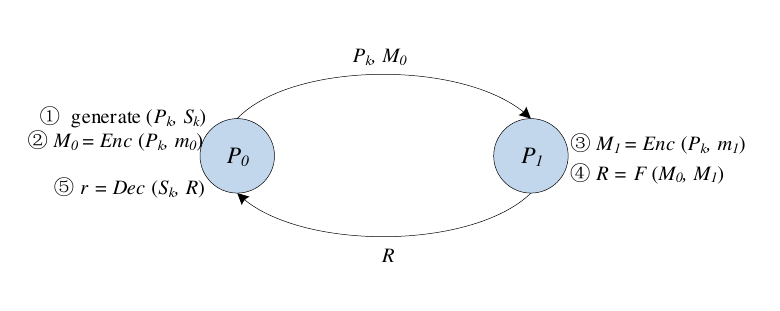}
    \caption{Two parties calculate a function with HE, where $(P_k,S_k)$ is a key-pair generated by $P_0$; $m_1$, $M_0$ and $m_1$, $M_1$ is the raw data and encrypted data of $P_0$ and $P_1$, respectively; r and R is the unencrypted and encrypted result, respectively.}
    \label{fig.HE}
\end{figure}

By performing calculations on encrypted data directly, HE can be used to ensure the security of the computing process.
HE can be adopted to protect data privacy through calculating and communicating on ciphertexts directly when training machine learning models.
As shown in Figure~\ref{fig.HE}, if we take secure two-party computation as an example, one party $P_0$ generates a key-pair $(P_k,S_k)$ for the homomorphic cryptosystem and sends the public key $P_k$ together with his encrypted message $M_0$ to the other party $P_1$, who then evaluate the Arithmetic circuit under encryption using the homomorphic properties of the cryptosystem with $M_0$.  
Finally, $P_1$ sends back the encrypted result $R$ which $P_0$ can decrypt using his private key $S_k$. 
Thus, the unencrypted data themselves are not transmitted. Nor can they be guessed by other parties. Note that there is only a little possibility of leaking the original data.  
Moreover, HE methods are widely applied in MPL frameworks to generate multiplication triples in the pre-computation phase~\cite{mohassel2017secureml}, which can efficiently reduce the communication overhead in the online phase. 
Note that non-linear functionalities such as the ReLU, and Sigmod activation functions cannot be supported directly by PHE or SWHE. 
 
Wu et al.~\cite{wu2013privacy} trained a privacy-preserving LOR model using AHE on vertically partitioned data. However, the time complexity of~\cite{wu2013privacy} increases exponentially according to the number of parameters.
Giacomelli et al.~\cite{giacomelli2018privacy} proposed an approach that trained a linear regression (LR for short) model using only HE in both the horizontally partitioned setting and vertically partitioned setting. 
Besides, CryptoNets~\cite{gilad2016cryptonets}, CryptoDL~\cite{hesamifard2017cryptodl} and SEALion~\cite{van2019sealion} are HE-based frameworks for the inference based on convolutional neural networks (CNNs for short), but they cannot support  the training.

\smallskip
\noindent\textbf{Features:} \emph{HE-based MPL frameworks have low communication overhead with a constant number of communication rounds. However, these frameworks usually have enormous computational complexity due to expensive public-key operations in the online phase.}
\smallskip

\subsubsection{GC-based MPL Frameworks}


To the best of our knowledge, the first implementation of GC protocol, which supported evaluating generic function securely between two parties, was the Fairplay framework~\cite{malkhi2004fairplay} in the semi-honest security model.
Fairplay allowed developers to specify the function to be computed securely with a high-level language, SFDL, which compiled and optimized into a Boolean circuit stored in a file. 
FairplayMP~\cite{ben2008fairplaymp} extended the original Fairplay framework to multiple parties. 
Mohassel et al.~\cite{mohassel2015fast} introduced a three-party GC-based protocol for a malicious security model.
TinyGarble~\cite{songhori2015tinygarble} used a sequential circuit description to generate compact and efficient Boolean circuits.
ObliVM~\cite{liu2015oblivm} offered a programming framework for secure computation.
The study in~\cite{gascon2017privacy} designed an MPL framework based on GC to train an LR model over vertically partitioned data distributed among an arbitrary number of parties.
DeepSecure~\cite{rouhani2018deepsecure} was designed based on GC protocol but only supported deep learning inference.

\smallskip
\noindent\textbf{Features:} \emph{GC-based MPL frameworks are mostly used in the scenario of two-party. 
These frameworks have a constant number of communication rounds, but the communication overhead is expensive, proportional to the size of the circuit.}    
\smallskip

\subsubsection{SS-based MPL Frameworks}

In SS-based MPL frameworks, parties initially share their inputs using a SS scheme. And throughout the training process, all data held by each party are in the form of shares.
To the best of our knowledge, Choi et al.~\cite{choi2012secure} first implemented the GMW protocol for any number of parties in the semi-honest security model.
A runtime environment for executing secure programs via the SPDZ protocol in the pre-processing model was presented in~\cite{keller2013architecture}. 
Cramer et al. presented SPD$\mathbb{Z}_{2^k}$~\cite{DBLP:conf/crypto/CramerDESX18} for a malicious security model with a dishonest majority that worked over rings instead of fields. Damg{\aa}rd et al.~\cite{damgaard2019new} optimized~\cite{DBLP:conf/crypto/CramerDESX18} and implemented their protocols to FRESCO~\cite{fresco} framework to support private classification using decision trees and support vector machines (SVMs). 
Makri et al. proposed EPIC~\cite{makri2017pics}\cite{makri2019epic}, which was secure against malicious adversaries and involved more than two parties. They used SPDZ to implement private image classification based on SVM. 
But both~\cite{damgaard2019new} and~\cite{makri2019epic} only focused on secure inference.
SecureNN~\cite{wagh2019securenn} provided a three-party method that eliminated expensive cryptographic operations for deep neural networks (DNN for short) and CNN training and inference with both semi-honest and malicious security models. 
FLASH~\cite{byali2020flash} proposed a four-party MPL framework for machine learning models, such as LR, LOR, DNN and binarized neural network (BNN for short).
Wagh et al. presented an efficient three-party MPL framework Falcon~\cite{wagh2020falcon} that utilized SS technologies for secure training and inference of NNs in an honest-majority setting. It designed an efficient protocol for batch normalization, which is critical for deep neural network training.
Koti et al. designed an SS-based MPL framework SWIFT~\cite{koti2021swift},  the key of which lay a maliciously secure, three-party computation over rings in the honest-majority setting, and they extended it to four parties. SWIFT implemented secure training and secure inference for LOR and secure inference for DNN.
Fantastic Four~\cite{dalskov2021fantastic} introduced a four-party honest-majority protocol with a malicious security model.
Song et al.~\cite{DBLP:conf/ccs/SongWWTLRWH22} proposed a robust MPL framework
with a privileged party pMPL. It guarantees: (i) Only the privileged party can get the final trained model, even if two assistant parties collude with each
other; (ii) The training process could be continued to the end if one assistant party drops out. pMPL is the first MPL framework that follows a hierarchical structure.

\smallskip
\noindent\textbf{Features:} \emph{SS-based MPL frameworks are  suitable for multiple parties in general. Commonly used SS protocol in SS-based MPL frameworks including GMW, BGW, RSS, SPDZ.}
\smallskip


\subsubsection{Mixed-Protocol based MPL Frameworks}
In addition to the aforementioned single-protocol MPL frameworks, a few commonly used frameworks typically adopt multiple protocols to leverage their advantages.
For instance, the essential idea behind the mixed protocol that combines HE and GC is to calculate operations that have an efficient representation as Arithmetic circuits (e.g., additions and multiplications) using HE and 
operations that have an efficient representation as Boolean circuits (e.g., comparisons) using GC.
However, the conversion between different schemes of shares is not trivial and the costs are relatively expensive.

The TASTY compiler~\cite{henecka2010tasty} could automatically generate protocols based on HE and GC as well as combinations of both.
The study of ~\cite{nikolaenko2013privacy-preserving} combined HE and GC to learn an LR model with horizontally partitioned data.
Demmler et al. proposed ABY~\cite{demmler2015aby}, a mixed-protocol framework combining multiple MPC protocols for two-party computation under the semi-honest security model. 
ABY3~\cite{mohassel2018aby3} extended ABY to three-party scenarios.
Trident~\cite{DBLP:conf/ndss/ChaudhariRS20} outperformed ABY3 in terms of communication complexity with the privilege of having an extra honest party and extended ABY3 to four parties.
ABY2.0~\cite{patra2021aby2} optimized the ABY framework and greatly improved its performance.
Mohassel et al.~\cite{mohassel2017secureml} proposed SecureML, which combined additive SS and GC in the online phase. It focused on both secure training and inference of various machine learning models like LR, LOR and NN.
QUOTIENT~\cite{agrawal2019quotient} implemented SS and Yao's GC to train DNN models with two parties in the semi-honest security model with an honest majority setting.
BLAZE~\cite{DBLP:conf/ndss/PatraS20} combined GC and SS to perform secure training and secure inference for LR and LOR models and secure inference alone for DNN.
CAESAR~\cite{chen2020homomorphic} combined HE and SS to build secure large-scale sparse LOR model in the vertically partitioned setting between two parties.
Tan et al. designed CRYPTGPU~\cite{tan2021cryptgpu}, a three-party MPL framework that implements all of the cryptographic operations on the GPU in the semi-honest honest majority security model. It was built on top of PyTorch~\cite{paszke2019pytorch} and CRYPTEN~\cite{knott2021crypten} and supported more complex NN models as well as handle larger datasets, such as ImageNet.
What is more, several Mixed-protocol based MPL Frameworks were proposed only for secure inference, such as MiniONN~\cite{liu2017oblivious}, GAZELLE~\cite{juvekar2018gazelle}, Chameleon~\cite{riazi2018chameleon}, XONN~\cite{riazi2019xonn}, ASTRA~\cite{chaudhari2019astra}, Delphi~\cite{mishra2020delphi}, CrypTFlow2~\cite{rathee2020cryptflow2}.

\smallskip
\noindent\textbf{Features:}  \emph{Mixed-Protocol based MPL frameworks make a trade-off between computational complexity and communication overhead and make good use of the characteristics and advantages of different protocols.  However, there are relatively expensive conversion costs between different protocols.}
\smallskip

\subsection{Federated Learning}
\label{sec.fl}
Although the local data held by parties are not exchanged in FL frameworks, the \lgp{model updates} transmitted between the server and parties might leak sensitive information~\cite{aono2017privacy}\cite{melis2019exploiting}\cite{zhu2019deep}\cite{li2019quantification}.
To protect the parties' local data from being leaked, a few privacy-enhancing technologies are applied in FL frameworks to privately exchange \lgp{model updates} when parties interact with the server.
In this paper, we classify FL frameworks as Non-cryptographic FL frameworks, DP-based FL frameworks, and SC-based FL frameworks in terms of the privacy protection mechanisms used in FL frameworks. 

\subsubsection{Non-cryptographic FL frameworks}
Lots of FL frameworks focus on improving efficiency or meeting statistical challenges, while ignoring the potential risk brought by exchanging plaintext \lgp{model updates}, i.e. following the original paradigm proposed by Google~\cite{mcmahan2016federated}.

Smith et al. proposed a systems-aware optimization framework named MOCHA~\cite{smith2017federated}, which combined FL with multi-task learning to handle the statistical challenges.
FedCS~\cite{nishio2019client}, a mobile edge computing framework for machine learning presented by Nishio et al., aims to perform FL efficiently under the setting of heterogeneous parties.
Liang et al. proposed LG-FEDAVG~\cite{liang2020think} by combing local representation learning with FL. They showed that local models can better deal with statistical challenges and effectively learn fair representations that obfuscate protected attributes.    
Kone{\v{c}}n{\`y} et al. proposed FSVRG~\cite{konevcny2016federated1} to train a high-quality centralized model. Liu et al. designed the FedBCD~\cite{liu2019communication} algorithm, in which each party conducts 
multiple local updates before each communication to effectively reduce the number of communication rounds among parties. CFL~\cite{sattler2020clustered} presented by Sattler et al. utilizes geometric properties of the FL loss surface to group the client population into clusters with jointly trainable data distributions. Nori et al. proposed an enhanced FL scheme, namely FFL~\cite{nori2021fast}, which jointly and dynamically adjusts two variables: local update coefficients, and sparsity budgets of gradient compression, to minimize the learning error.

\smallskip
\noindent\textbf{Features:}  \emph{No additional security is provided for \lgp{model updates}, which are exchanged in plaintext in Non-cryptographic FL frameworks. Non-cryptographic FL frameworks are more efficient than the other two types of FL frameworks.} 
\smallskip

\subsubsection{SC-based FL frameworks}
Cryptographic primitives and protocols, such as HE, GC and SS, are widely used in FL frameworks.
In fact, the ways HE is applied to FL frameworks are similar to that applied to MPL frameworks except for some details.  
In FL frameworks, HE is used to protect the privacy of the \lgp{model updates} interacted between the party and the server rather than the data interacted between parties as HE applied in MPL frameworks. The study~\cite{aono2017privacy} applied AHE to preserve the privacy of gradients for providing security against the semi-honest and centralized server in FL models.

In addition, HE and SS protocols involve multiple parties and retain the original accuracy with very high security. It guarantees each party knows nothing except the results and can be applied in FL models for secure aggregation and to protect local models. In the FL frameworks based on HE or SS protocols, the centralized server cannot obtain any local information and local updates, but observe the exact aggregated results at each round. However, HE or SS protocols applied in FL frameworks will incur significant extra communication and computation costs. At present, SS protocols, especially Shamir's SS protocol, as a basis are the most widely used in FL frameworks.
Bonawitz et al.~\cite{bonawitz2017practical} presented a protocol based on Shamir's SS for securely aggregating updates.
Cheng et al. proposed a novel lossless privacy-preserving tree-boosting system known as SecureBoost~\cite{cheng2019secureboost} in the setting of FL; Xu et al. presented VerifyNet~\cite{xu2019verifynet}, the first privacy-preserving and verifiable FL framework; Sav et al. designed a novel system, POSEIDON~\cite{sav2020poseidon}, to address the problem of privacy-preserving training and evaluation of neural networks in an N-party,  FL setting.

\smallskip
\noindent\textbf{Features:}  \emph{SC-based FL frameworks, compared with Non-cryptographic FL frameworks, have a lossless accuracy with higher security. However, these frameworks introduce significant extra communication or computational cost.}
\smallskip

\subsubsection{DP-based FL frameworks}
\label{sec.dp}

By adding random noises to the raw data or the \lgp{model updates}, DP achieves that the corrupted parties (or the centralized server) cannot infer any specific data owners' data are used in training from the obtained model updates.

\begin{figure}[htb]
\centering
\subfigure[Central Differential Privacy (CDP)]{
\label{fig.DP_central}
\includegraphics[scale=0.7]{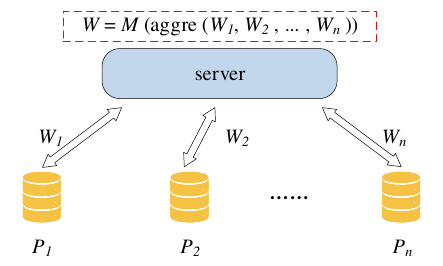}
}
\subfigure[Local Differential Privacy (LDP)]{
\label{fig.DP_local}
\includegraphics[scale=0.7]{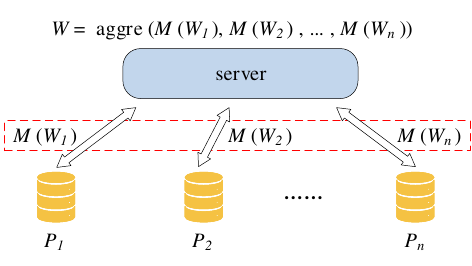}
}
\centering
\caption{CDP and LDP. Here, $M$ denotes the DP mechanism. $W_i$ denotes the model updates for each party $P_i$. With CDP (a), the model updates are aggregated by the centralized server (must be honest) and then perturb them. With LDP (b), the model updates are applied DP mechanisms and then aggregated by the centralized server (maybe dishonest). }
\label{fig.DP}
\end{figure}

In FL frameworks, the applications of DP can be classified into two categories: central DP (CDP for short) and local DP (LDP for short). 
\lgp{As shown in Figure \ref{fig.DP}, in the setting of CDP, the model updates uploaded by each party in each round are firstly aggregated by the centralized server. Then the noises are added to the aggregated global model updates. The noise will perturb the global model, thereby preventing any party from reconstructing the raw data of others by exploiting the global model to a certain extent.
In the setting of LDP, each party adds noise locally on model updates prior to uploading them to the centralized server, which can prevent the server from reconstructing the raw data of parties by exploiting the model updates.
Note that the centralized server generally must be trusted in the setting of CDP, while the centralized server can be dishonest in the setting of LDP.}
LDP affords stronger privacy than CDP, but achieving it while maintaining utility is a challenge~\cite{bassily2017practical}.
Additionally, there exists a trade-off between privacy and model accuracy, as adding more noise will provide a greater privacy guarantee but may compromise the model accuracy significantly.  In addition, k-anonymity\cite{sweeney2002k}\cite{el2008protecting}, l-diversity~\cite{machanavajjhala2007diversity}\cite{zhou2011k} and t-closeness~\cite{li2007t} can also apply to FL for data privacy protection.


The study~\cite{geyer2017differentially} applied CDP to FL frameworks which can maintain client-level DP by hiding the parties' contributions during the training and made a trade-off between privacy and performance.
McMahan et al.~\cite{mcmahan2017learning} used a similar method to train Long Short-Term Memory (LSTM) models.
LDP-FedSGD~\cite{zhao2020local} presented by Zhao et al. integrates FL and LDP to facilitate the crowdsourcing applications to achieve the machine learning model; Liu et al. first proposed an FL framework FLAME~\cite{liu2020flame} in the shuffle model; Li et al.~\cite{li2021rr}  designed a privacy-enhanced FL scheme leveraging randomized response mechanism and the local adaptive differential privacy mechanism. Sun et al.~\cite{sun2020ldp}  proposed a novel design of an LDP for FL.

\smallskip
\noindent\textbf{Features:}  \emph{DP-based FL frameworks provide privacy guarantees by adding noise to the \lgp{model updates} or raw data. 
Compared with SC-based FL frameworks, these frameworks have cheap computational and communication costs but need a compromise of accuracy.}
\smallskip


\subsection{Comparisons}
\label{sec.compar}


\subsubsection{Comparisons of two technical folders}

\begin{itemize}
    \item According to the insights in Section \ref{sec.challenge}, \emph{MPL frameworks usually provide higher security than FL frameworks, but FL frameworks are more efficient. In addition, the data distributions (IID or non-IID) have no effect on MPL frameworks.}
    \item \emph{The Accuracy of trained models.} In MPL frameworks, there is usually no loss of accuracy in the global model. But FL frameworks generally have a certain loss of accuracy, especially in the non-IID setting.
    \item \emph{MPL and FL are suitable for different scenarios.} Combined with the above analysis, we can find that MPL is more suitable for scenarios with higher security or complex data distributions, and FL is more suitable for scenarios with the requirement of the higher performance \hwl{of training with} more parties. 
\end{itemize}



\begin{table*}[]
\centering
\caption{Comparisons among eighteen MPL frameworks. Note that (1) SH, Mal, HM, DM refer to semi-honest, malicious, honest-majority, and dishonest-majority security models, respectively; (2) 3-layer DNN refers to a network structure with three fully connected layers, 3-layer CNN refers to a network structure with one convolutional layer and two fully connected layers, 4-layer CNN refers to a network structure with two convolutional layers and two fully connected Layer network structure; (3) the model with (*) indicates that the framework only supports the inference of this model. (4)The accuracy of 3-layer CNN in ABY3 is estimated according to the plaintext model.}
\label{table.frameworks}
\renewcommand\arraystretch{1}
\begin{tabular}{|c|c|c|c|c|c|c|c|c|}
\hline
\multirow{2}{*}{Framework}                 & \multirow{2}{*}{Technical routes}                                                                  & \multirow{2}{*}{\begin{tabular}[c]{@{}c@{}}Parties \\ supported\end{tabular}} & \multirow{2}{*}{\begin{tabular}[c]{@{}c@{}}Security\\ model\end{tabular}}   & \multirow{2}{*}{\begin{tabular}[c]{@{}c@{}}Data \\ partitioning\end{tabular}} & \multirow{2}{*}{\begin{tabular}[c]{@{}c@{}}Models \\ supported\end{tabular}} & \multirow{2}{*}{\begin{tabular}[c]{@{}c@{}}Activation \\ function\end{tabular}} & \multirow{2}{*}{NN accuracy}                                                                                                    & \multirow{2}{*}{Year} \\
                                           &                                                                                                    &                                                                               &                                                                             &                                                                               &                                                                              &                                                                                 &                                                                                                                                 &                       \\ \hline
\cite{wu2013privacy}                       & \multirow{2}{*}{\begin{tabular}[c]{@{}c@{}}HE-based MPL \\ framework\end{tabular}}                 & 2                                                                             & $\backslash$                                                                & Vertical                                                                      & LOR                                                                          & $\backslash$                                                                    & $\backslash$                                                                                                                    & 2013                  \\ \cline{1-1} \cline{3-9} 
\cite{giacomelli2018privacy}               &                                                                                                    & 2                                                                             & SH                                                                          & \begin{tabular}[c]{@{}c@{}}Horizontal\\ Vertical\end{tabular}                 & LR                                                                           & $\backslash$                                                                    & $\backslash$                                                                                                                    & 2018                  \\ \hline
\cite{gascon2017privacy}                   & \begin{tabular}[c]{@{}c@{}}GC-based MPL \\ framework\end{tabular}                                  & 2+                                                                            & SH                                                                          & Vertical                                                                      & LR                                                                           & $\backslash$                                                                    & $\backslash$                                                                                                                    & 2017                  \\ \hline
SecureNN~\cite{wagh2019securenn}           & \multirow{5}{*}{\begin{tabular}[c]{@{}c@{}}SS-based MPL \\ framework\end{tabular}}                 & 3                                                                             & \begin{tabular}[c]{@{}c@{}}SH, HM\\ Mal, HM\end{tabular}                    & \begin{tabular}[c]{@{}c@{}}Horizontal\\ Vertical\end{tabular}                 & NN (DNN, CNN)                                                                & ReLU                                                                            & \begin{tabular}[c]{@{}c@{}}93.4\% (3-layer DNN)\\ 98.77\% (4-layer CNN)\\ 99.15\% (LeNet)\end{tabular}                          & 2019                  \\ \cline{1-1} \cline{3-9} 
FLASH~\cite{byali2020flash}                &                                                                                                    & 4                                                                             & \begin{tabular}[c]{@{}c@{}}SH, HM\\ Mal, HM\end{tabular}                    & \begin{tabular}[c]{@{}c@{}}Horizontal\\ Vertical\end{tabular}                 & \begin{tabular}[c]{@{}c@{}}LR, LOR,\\ NN (DNN, BNN)\end{tabular}             & \begin{tabular}[c]{@{}c@{}}ReLU\\ Sigmoid\end{tabular}                          & $\backslash$                                                                                                                    & 2020                  \\ \cline{1-1} \cline{3-9} 
Falcon~\cite{wagh2020falcon}               &                                                                                                    & 3                                                                             & \begin{tabular}[c]{@{}c@{}}SH, HM\\ Mal, HM\end{tabular}                    & \begin{tabular}[c]{@{}c@{}}Horizontal\\ Vertical\end{tabular}                 & NN (DNN, CNN)                                                                & ReLU                                                                            & \begin{tabular}[c]{@{}c@{}}97.42\% (3-layer DNN)\\ 97.81\% (3-layer CNN)\\ 98.64\% (4-layer CNN)\\ 99.15\% (LeNet)\end{tabular} & 2020                  \\ \cline{1-1} \cline{3-9} 
SWIFT~\cite{koti2021swift}                 &                                                                                                    & 3,4                                                                           & Mal, HM                                                                      & \begin{tabular}[c]{@{}c@{}}Horizontal\\ Vertical\end{tabular}                 & \begin{tabular}[c]{@{}c@{}}LOR,\\ NN*(DNN)\end{tabular}                      & \begin{tabular}[c]{@{}c@{}}ReLU\\ Sigmoid\end{tabular}                          & $\backslash$                                                                                                                    & 2021                  \\ \cline{1-1} \cline{3-9} 
Fantastic Four~\cite{dalskov2021fantastic} &                                                                                                    & 3,4                                                                           & \begin{tabular}[c]{@{}c@{}}SH, HM\\ Mal, HM\end{tabular}                    & \begin{tabular}[c]{@{}c@{}}Horizontal\\ Vertical\end{tabular}                 & \begin{tabular}[c]{@{}c@{}}LOR,\\ NN (DNN)\end{tabular}                      & \begin{tabular}[c]{@{}c@{}}ReLU\\ Sigmoid\\ Softmax\end{tabular}                & \begin{tabular}[c]{@{}c@{}}92.3\% (1-layer DNN)\\ 95.0\% (2-layer DNN)\\ 92.9\% (3-layer DNN)\end{tabular}                      & 2021                  \\ \cline{1-1} \cline{3-9}
pMPL~\cite{DBLP:conf/ccs/SongWWTLRWH22}                 &                                                                                                    & 2,3                                                                           & SH                                                                      & \begin{tabular}[c]{@{}c@{}}Horizontal\\ Vertical\end{tabular}                 & \begin{tabular}[c]{@{}c@{}}LR, LOR,\\ NN*(DNN)\end{tabular}                      & \begin{tabular}[c]{@{}c@{}}ReLU\\ Sigmoid\end{tabular}                          & 96\% (3-layer DNN)                                                                                                                     & 2022                  \\ \hline 
\cite{nikolaenko2013privacy-preserving}    & \multirow{10}{*}{\begin{tabular}[c]{@{}c@{}}Mixed-protocol \\ based \\ MPL framework\end{tabular}} & 2+                                                                            & \begin{tabular}[c]{@{}c@{}}SH;\\ Mal\end{tabular}                            & Horizontal                                                                    & LR                                                                           & $\backslash$                                                                    & $\backslash$                                                                                                                    & 2013                  \\ \cline{1-1} \cline{3-9} 
SecureML~\cite{mohassel2017secureml}       &                                                                                                    & 2                                                                             & SH                                                                          & \begin{tabular}[c]{@{}c@{}}Horizontal\\ Vertical\end{tabular}                 & \begin{tabular}[c]{@{}c@{}}LR, LOR,\\ NN (DNN)\end{tabular}                  & \begin{tabular}[c]{@{}c@{}}ReLU\\ Sigmoid\\ Softmax\end{tabular}                & 93.4\% (3-layer DNN)                                                                                                            & 2017                  \\ \cline{1-1} \cline{3-9} 
ABY3~\cite{mohassel2018aby3}               &                                                                                                    & 3                                                                             & \begin{tabular}[c]{@{}c@{}}SH, HM\\ Mal, HM\end{tabular}                    & \begin{tabular}[c]{@{}c@{}}Horizontal\\ Vertical\end{tabular}                 & \begin{tabular}[c]{@{}c@{}}LR, LOR \\ \\ NN (DNN, CNN)\end{tabular}          & \begin{tabular}[c]{@{}c@{}}ReLU\\ Sigmoid\\ Softmax\end{tabular}                & \begin{tabular}[c]{@{}c@{}}94\% (3-layer DNN)\\ 99\% (3-layer CNN)\end{tabular}                                                 & 2018                  \\ \cline{1-1} \cline{3-9} 
QUOTIENT~\cite{agrawal2019quotient}        &                                                                                                    & 2                                                                             & SH                                                                          & \begin{tabular}[c]{@{}c@{}}Horizontal\\ Vertical\end{tabular}                 & NN (DNN, CNN)                                                                & \begin{tabular}[c]{@{}c@{}}ReLU\\ Sigmoid\\ Softmax\end{tabular}                & 99.38\% (3-layer DNN)                                                                                                           & 2019                  \\ \cline{1-1} \cline{3-9} 
Trident~\cite{DBLP:conf/ndss/ChaudhariRS20}          &                                                                                                    & 4                                                                             & \begin{tabular}[c]{@{}c@{}}SH, HM\\ Mal, HM\end{tabular}                    & \begin{tabular}[c]{@{}c@{}}Horizontal\\ Vertical\end{tabular}                 & \begin{tabular}[c]{@{}c@{}}LR, LOR, \\ NN (DNN, CNN)\end{tabular}            & \begin{tabular}[c]{@{}c@{}}ReLU\\ Sigmoid\end{tabular}                          & 98.3\% (3-layer CNN)                                                                                                            & 2020                  \\ \cline{1-1} \cline{3-9} 
BLAZE~\cite{DBLP:conf/ndss/PatraS20}                &                                                                                                    & 3                                                                             & \begin{tabular}[c]{@{}c@{}}SH, HM\\ Mal, HM\end{tabular}                    & \begin{tabular}[c]{@{}c@{}}Horizontal\\ Vertical\end{tabular}                 & \begin{tabular}[c]{@{}c@{}}LR, LOR,\\ NN* (DNN)\end{tabular}                 & \begin{tabular}[c]{@{}c@{}}ReLU\\ Sigmoid\end{tabular}                          & (97.8\% 3-layer DNN)                                                                                                            & 2020                  \\ \cline{1-1} \cline{3-9} 
MP-SPDZ~\cite{keller2020mp}                &                                                                                                    & 2+                                                                            & \begin{tabular}[c]{@{}c@{}}SH, HM\\ Mal, HM\\ SH, DM\\ Mal, DM\end{tabular} & \begin{tabular}[c]{@{}c@{}}Horizontal\\ Vertical\end{tabular}                 & $\backslash$                                                                 & $\backslash$                                                                    & $\backslash$                                                                                                                    & 2020                  \\ \cline{1-1} \cline{3-9} 
CRYPTGPU~\cite{tan2021cryptgpu}            &                                                                                                    & 2                                                                             & SH                                                                          & \begin{tabular}[c]{@{}c@{}}Horizontal\\ Vertical\end{tabular}                 & NN (CNN)                                                                     & ReLU                                                                            & 93.97\% (LeNet)                                                                                                                 & 2021                  \\ \cline{1-1} \cline{3-9} 
CAESAR~\cite{chen2020homomorphic}          &                                                                                                    & 2                                                                             & SH                                                                          & Vertical                                                                      & LOR                                                                          & $\backslash$                                                                    & $\backslash$                                                                                                                    & 2021                  \\ \cline{1-1} \cline{3-9} 
ABY2.0~\cite{patra2021aby2}                &                                                                                                    & 2                                                                             & SH                                                                          & \begin{tabular}[c]{@{}c@{}}Horizontal\\ Vertical\end{tabular}                 & \begin{tabular}[c]{@{}c@{}}LR,\\ NN (DNN, CNN),\\ SVM*, BNN*\end{tabular}    & \begin{tabular}[c]{@{}c@{}}ReLU\\ Sigmoid\end{tabular}                          & $\backslash$                                                                                                                    & 2021                  \\ \hline
\end{tabular}
\end{table*}

\begin{table*}[t]
    \caption{Comparisons among eighteen FL frameworks}
    \label{table.flframeworks}
    \centering
    \renewcommand\arraystretch{1}
\begin{tabular}{|c|c|c|c|c|c|c|}
\hline
                                                          & Technical routes                                                                           & Parties supported & \multicolumn{1}{l|}{Security model}                             & Data partitioning & Models supported                                               & Year                      \\ \hline
FedAvg~\cite{mcmahan2016federated}                        & \multirow{8}{*}{\begin{tabular}[c]{@{}c@{}}Non-cryptographic \\ FL framework\end{tabular}} & 2+                & $\backslash$                                                    & Horizontal        & \begin{tabular}[c]{@{}c@{}}NN (DNN, CNN), \\ LSTM\end{tabular} & 2016                      \\ \cline{1-1} \cline{3-7} 
FSVRG~\cite{konevcny2016federated1}                       &                                                                                            & 2+                & $\backslash$                                                    & Horizontal        & LR, LOR, SVM                                                    & 2016                      \\ \cline{1-1} \cline{3-7} 
MOCHA~\cite{smith2017federated}                           &                                                                                            & 2+                & $\backslash$                                                    & Horizontal        & $\backslash$                                                   & 2017                      \\ \cline{1-1} \cline{3-7} 
FedCS~\cite{nishio2019client}                             &                                                                                            & 2+                & $\backslash$                                                    & Horizontal        & NN (CNN)                                                       & 2018                      \\ \cline{1-1} \cline{3-7} 
LG-FEDAVG~\cite{liang2020think}                           &                                                                                            & 2+                & $\backslash$                                                    & Horizontal        & NN (CNN)                                                       & 2019                      \\ \cline{1-1} \cline{3-7} 
FedBCD~\cite{liu2019communication}                        &                                                                                            & 2+                & $\backslash$                                                    & Vertical          & NN (DNN)                                                       & 2019                      \\ \cline{1-1} \cline{3-7} 
CFL~\cite{sattler2020clustered}                           &                                                                                            & 2+                & $\backslash$                                                    & Horizontal        & NN (CNN), LSTM                                                 & 2021                      \\ \cline{1-1} \cline{3-7} 
FFL~\cite{nori2021fast}                                   &                                                                                            & 2+                & $\backslash$                                                    & Horizontal        & NN (CNN, DNN)                                                  & 2021                      \\ \hline
Federated Secure Aggregation~\cite{bonawitz2017practical} & \multirow{4}{*}{\begin{tabular}[c]{@{}c@{}}SC-based \\ FL framework\end{tabular}}          & 2+                & \begin{tabular}[c]{@{}c@{}}Semi-honest\\ Malicious\end{tabular} & Horizontal        & NN (DNN)                                                       & 2017                      \\ \cline{1-1} \cline{3-7} 
SecureBoost~\cite{cheng2019secureboost}                   &                                                                                            & 2+                & Semi-honest                                                     & Vertical          & Decision Trees                                                 & 2019                      \\ \cline{1-1} \cline{3-7} 
VerifyNet~\cite{xu2019verifynet}                          &                                                                                            & 2+                & Semi-honest                                                     & Horizontal        & NN (CNN)                                                       & 2019                      \\ \cline{1-1} \cline{3-7} 
POSEIDON~\cite{sav2020poseidon}                           &                                                                                            & 2+                & Semi-honest                                                     & Horizontal        & NN                                                             & 2021                      \\ \hline
\cite{geyer2017differentially}                            & \multirow{6}{*}{\begin{tabular}[c]{@{}c@{}}DP-based \\ FL framework\end{tabular}}          & 2+                & $\backslash$                                                    & Horizontal        & $\backslash$                                                   & \multicolumn{1}{l|}{2017} \\ \cline{1-1} \cline{3-7} 
DP-FedAvg~\cite{mcmahan2017learning}                      &                                                                                            & 2+                & $\backslash$                                                    & Horizontal        & LSTM                                                           & 2018                      \\ \cline{1-1} \cline{3-7} 
LDP-FedSGD~\cite{zhao2020local}                           &                                                                                            & 2+                & $\backslash$                                                    & Horizontal        & LR, LOR, SVM                                                   & 2020                      \\ \cline{1-1} \cline{3-7} 
FLAME~\cite{liu2020flame}                                 &                                                                                            & 2+                & $\backslash$                                                    & Horizontal        & LOR                                                            & 2021                      \\ \cline{1-1} \cline{3-7} 
RR-LADP~\cite{li2021rr}                                   &                                                                                            & 2+                & $\backslash$                                                    & Horizontal        & NN (DNN)                                                       & 2021                      \\ \cline{1-1} \cline{3-7} 
LDP-FL~\cite{sun2020ldp}                                                    &                                                                                            & 2+                & $\backslash$                                                    & Horizontal        & NN(CNN)                                                        & \multicolumn{1}{l|}{2021} \\ \hline
\end{tabular}
\end{table*}



\subsubsection{Comparisons of frameworks}
\label{sec.comparison}
As shown in Table \ref{table.frameworks} and Table \ref{table.flframeworks}, we investigate eighteen MPL frameworks and eighteen FL frameworks. Then we group these MPL frameworks into four technical routes according to the description in Section \ref{sec.mpc}, as well as grouping the FL frameworks into three technical routes.  In addition, we list the information of these frameworks in items of the number of parties supported, security models implemented, the ways data are partitioned, and the types of machine learning models supported. All of these frameworks implement multiple parties training machine learning models with privacy preservation. 
Finally, we can conclude as follows:

\begin{itemize}

    \item The major studied frameworks in the solutions of MPL are the SS-based MPL frameworks and the Mixed-protocol based MPL frameworks. We argue the reason is that HE-based MPL frameworks have immense computational complexity, and the GC-based MPL frameworks have enormous communication overhead, both of which could be unacceptable in \hwl{efficient} MPL frameworks.
    SS-based MPL frameworks and Mixed-protocol MPL frameworks make a trade-off between communication overhead and computational complexity, suitable for more practical scenarios. 
    
    \item The major studied frameworks in the solutions of FL are Non-cryptographic FL frameworks.  These frameworks are mainly designed to improve the efficiency in scenarios such as non-IID distribution settings or vertically partitioned settings.
    However, to improve the security of FL frameworks, the studies about SC-based FL frameworks and \lgp{HE}-based FL frameworks are also significant and should be studied deeply.
    
    
    \item The current MPL frameworks usually use tailored protocols to support a specific number of parties, that is, poor scalability. In addition, the participating parties involved in the current framework are generally two or three parties; when there are more parties, even dozens of them, the efficiency of the framework drops sharply. However, FL frameworks generally supported multiple, e.g. tens or hundreds of, parties.
    
    
    \item The existing FL frameworks usually consider a scenario where the data among parties are horizontally partitioned, while only a few works consider the vertically partitioned data. We argue that the processing of training data and the design of the framework in horizontal data partitioning is relatively easier than vertical data partitioning.
    However MPL frameworks can easily support the scenario of vertical data partitioning, after each party shares its own data with others, the training does not make full use of the characteristics of data vertically partitioned. Considering the fact that vertical data partitioning is also expected and essential in the real world, it remains promising to train complex machine learning models efficiently upon vertical data partitioning.
    
    \item Most of the current frameworks are  targeted at training parametric models, such as LR, LOR, NN (including DNN, CNN), while a few frameworks are designed to train non-parametric models such as decision trees. 
\end{itemize}

\section{Platforms}
\label{sec:platforms}
In recent years, many open-source platforms related to \TMMPP have been developed by international and well-known business enterprises. 
Due to MPL platforms generally involve one technical route, for instance, \texttt{\lgp{Queqiao}}\footnote{\url{https://github.com/FudanMPL/SecMML}} only based on BGW, but FL platforms are usually designed to be more complex and suitable for a variety of scenarios.
So, in this section, we focus on five popular FL platforms (\texttt{FATE}~\footnote{\url{https://github.com/FederatedAI/FATE}}, \texttt{TFF}~\footnote{\url{https://github.com/tensorflow/federated}}, \texttt{PaddleFL}\footnote{\url{https://github.com/PaddlePaddle/PaddleFL}}, \texttt{Pysyft}~\footnote{\url{https://github.com/OpenMined/PySyft}}, \texttt{coMind})\footnote{\url{https://github.com/coMindOrg/federated-averaging-tutorials}}, and present comparisons among them according to basic information and experiments. 



\begin{table*}[]
    \caption{comparisons among five open-source platforms}
    \label{open-source}
    \centering
    \renewcommand\arraystretch{1}
    \begin{tabular}{|c|c|c|c|c|c|c|}
    \hline
    \multicolumn{2}{|c|}{Indicator/Platform}              & \texttt{FATE}  & \texttt{TFF}  & \texttt{PaddleFL}  & \texttt{Pysyft}  & \texttt{coMind}  \\ \hline
    \multirow{3}{*}{Data Partitioning}  & Horizontal        & \checkmark     & \checkmark    & \checkmark         & \checkmark       & \checkmark       \\ \cline{2-7} 
                                        & Vertical          & \checkmark     & $\times$      & \checkmark         & $\times$         & $\times$         \\ \cline{2-7} 
                                        & Transfer Learning & \checkmark     & $\times$      & \checkmark         & $\times$         & $\times$         \\ \hline
    \multirow{3}{*}{Privacy Technology} & HE                & \checkmark     & $\times$      &  $\times$          & \checkmark       & $\times$         \\ \cline{2-7} 
                                        & MPC               & \checkmark     & $\times$      & \checkmark         & \checkmark       & $\times$         \\ \cline{2-7} 
                                        & DP                & $\times$       & $\times$      & \checkmark         & \checkmark       & $\times$         \\ \hline
    \multirow{2}{*}{Way of Deployment} & Stand-alone       & \checkmark     & \checkmark    & \checkmark         & \checkmark       & \checkmark       \\ \cline{2-7} 
                                        & Cluster           & \checkmark     & $\times$      & \checkmark         & $\times$        & \checkmark         \\ \hline
    \multirow{2}{*}{Hardware}           & CPU               & \checkmark     & \checkmark    & \checkmark         & \checkmark       & \checkmark       \\ \cline{2-7} 
                                        & GPU               & $\times$       & \checkmark    & $\times$           & $\times$         & \checkmark       \\ \hline
    \multicolumn{2}{|c|}{Visual interface}                     & \checkmark     & \checkmark      & $\times$         & $\times$         & $\times$         \\ \hline

    \end{tabular}
\end{table*}

\subsection{Basic Information}
Table \ref{open-source} shows the information on these open-source platforms.
Overall, all \lgp{the} five platforms support horizontal data partitioning. \texttt{FATE} and \texttt{PaddleFL} also support vertical data partitioning and transfer learning. For security purposes, \texttt{FATE} employs two privacy technologies, i.e HE and MPC, and \texttt{PaddleFL} employs MPC and DP. \texttt{Pysyft} provides all these three privacy technologies, while \texttt{TFF} and \texttt{coMind} do not employ any privacy technologies. 
\texttt{TFF} and \texttt{Pysyft}  simulate the process of federated learning through multiple processes instead of directly supporting cluster deployment. To improve efficiency, \texttt{TFF} and \texttt{coMind} support GPU. Furthermore, \texttt{FATE} and \texttt{TFF} provide visual interfaces, which makes them more user-friendly.

\subsection{Experimental settings}
We have Linux servers equipped with 20-core 2.4 GHz Intel Xeon CPU and 128GB of RAM and use it to deploy multiple processes to simulate the training process.
In our experiments, we use the MNIST dataset\footnote{MNIST Dataset: http://yann.lecun.com/exdb/mnist/}.
It contains images of handwritten digits from “0” to “9”, each with 784 features representing 28 × 28 pixels in the image.
The training set contains 60,000 samples, and the test set contains 10,000 samples.
Under different numbers of parties, the training set is divided equally. For example, if there are 2 parties, each party has 30,000 training data, and 10 parties, each party has 6000 training data.

Among the five platforms, \texttt{coMind} and \texttt{PaddleFL} use the gRPC protocol to communicate, and they cannot run when there are more than 20 parties.  We argue this might be related to the allocation of system resources. \texttt{Fate} (standalone-1.4.0 version) cannot run when there are more than 50 parties for the reason that the filename length of the file storing the global model (associated with the number of parties) exceeds the Linux limitation.

\begin{table}[ht]
\caption{LOR and NN comparisons of accuracy and time of open source platforms for different numbers of parties }
\label{exper_lor}
\centering
\renewcommand\arraystretch{1}
\scalebox{0.9}{
\begin{tabular}{|c|c|cc|cc|}
\hline
\multirow{2}{*}{Platform} & \multirow{2}{*}{\begin{tabular}[c]{@{}c@{}}Number of \\ parties\end{tabular}} & \multicolumn{2}{c|}{LOR}                 & \multicolumn{2}{c|}{NN}                  \\ \cline{3-6} 
                          &                                                                               & \multicolumn{1}{l|}{Accuracy} & Times/s  & \multicolumn{1}{l|}{Accuracy} & Times/s  \\ \hline
\multirow{3}{*}{FATE}     & 2                                                                             & \multicolumn{1}{c|}{91.51\%}  & 262.701  & \multicolumn{1}{c|}{95.65\%}  & 281.728  \\ \cline{2-6} 
                          & 10                                                                            & \multicolumn{1}{c|}{89.23\%}  & 84.478   & \multicolumn{1}{c|}{91.49\%}  & 94.226   \\ \cline{2-6} 
                          & 20                                                                            & \multicolumn{1}{c|}{87.83\%}  & 76.846   & \multicolumn{1}{c|}{89.71\%}  & 87.837   \\ \hline
\multirow{5}{*}{TFF}      & 2                                                                             & \multicolumn{1}{c|}{91.57\%}  & 70.465   & \multicolumn{1}{c|}{95.11\%}  & 74.914   \\ \cline{2-6} 
                          & 10                                                                            & \multicolumn{1}{c|}{89.83\%}  & 61.468   & \multicolumn{1}{c|}{92.25\%}  & 66.686   \\ \cline{2-6} 
                          & 20                                                                            & \multicolumn{1}{c|}{88.46\%}  & 86.344   & \multicolumn{1}{c|}{90.69\%}  & 93.695   \\ \cline{2-6} 
                          & 50                                                                            & \multicolumn{1}{c|}{86.25\%}  & 173.598  & \multicolumn{1}{c|}{88.13\%}  & 188.620  \\ \cline{2-6} 
                          & 100                                                                           & \multicolumn{1}{c|}{82.81\%}  & 364.447  & \multicolumn{1}{c|}{84.66\%}  & 398.911  \\ \hline
\multirow{2}{*}{PaddleFL} & 2                                                                             & \multicolumn{1}{c|}{91.85\%}  & 521.415  & \multicolumn{1}{c|}{95.75\%}  & 663.270  \\ \cline{2-6} 
                          & 10                                                                            & \multicolumn{1}{c|}{89.96\%}  & 329.521  & \multicolumn{1}{c|}{92.41\%}  & 361.409  \\ \hline
\multirow{5}{*}{Pysyft}   & 2                                                                             & \multicolumn{1}{c|}{90.58\%}  & 2681.517 & \multicolumn{1}{c|}{92.65\%}  & 3612.837 \\ \cline{2-6} 
                          & 10                                                                            & \multicolumn{1}{c|}{90.67\%}  & 2645.998 & \multicolumn{1}{c|}{92.75\%}  & 3604.416 \\ \cline{2-6} 
                          & 20                                                                            & \multicolumn{1}{c|}{90.58\%}  & 2729.887 & \multicolumn{1}{c|}{85.58\%}  & 3655.997 \\ \cline{2-6} 
                          & 50                                                                            & \multicolumn{1}{c|}{90.71\%}  & 2715.742 & \multicolumn{1}{c|}{93.01\%}  & 3711.259 \\ \cline{2-6} 
                          & 100                                                                           & \multicolumn{1}{c|}{90.83\%}  & 2754.873 & \multicolumn{1}{c|}{92.89\%}  & 3882.532 \\ \hline
\multirow{2}{*}{coMind}   & 2                                                                             & \multicolumn{1}{c|}{95.31\%}  & 123.134  & \multicolumn{1}{c|}{96.88\%}  & 185.324  \\ \cline{2-6} 
                          & 10                                                                            & \multicolumn{1}{c|}{93.75\%}  & 21.111   & \multicolumn{1}{c|}{92.19\%}  & 44.031   \\ \hline
\end{tabular}}
\end{table}

\subsection{Experiments for LOR and NN}

We focus on LOR and NN in our conducted experiments. Here, NN consists of an input layer, two hidden layers, and an output layer.  We set the batch size to 64, the global learning rate to 1, the local learning rate to 0.01, the number of global iterations to 50, and one round of local iteration is performed per global iteration. 
We adopt Softmax as the activation function of LOR, ReLU and Softmax as the activation function of NN, and FedAvg as  the aggregation algorithm. 
Additionally, we use the Cross-entropy loss function and the SGD method to train the model. 
The experimental results are shown in Table~\ref{exper_lor}. 

From Table~\ref{exper_lor}, 
we can \hwl{conclude} that, among the five platforms, as the number of parties increases, the accuracy of the model shows a downward trend. In addition, the running time decreases first, then increases \lgp{(except for \texttt{Pysyft})}. \lgp{The downward trend of accuracy is because, as the number of parties increases, the difference between/among the data distribution of the parties becomes more significant. And a greater difference in data distribution usually causes a lower model accuracy in FL. It is unavoidable that the data distribution of the parties becomes different as the number of parties increases, even if we randomly allocate all the data evenly to all parties.  When considering an extreme case, we randomly allocate 60,000 samples to 60,000 parties on average, and each participant has only one sample. As far as the label is concerned, there are ten completely different distributions among the parties. In addition, the reason why the running time decreases first and then increases is as follows.}
When there are only two parties, the low parallelism results in low CPU resource utilization. As the number of parties increases, the parallelism increases, and the CPU resource utilization increases. However, when the number of parties increases to a certain number, the CPU resource utilization reaches its peak. At this time, adding more parties (more processes) will increase memory consumption and thus increase the CPU switching time, so the running time increases. \lgp{In addition, the running time of \texttt{Pysyft} is much longer than the other four platforms, and there is no noticeable change with the increase of parties. This 
might be because the underlying implementations of \texttt{Pysyft} are different from other frameworks: (1) The parties in \texttt{Pysyft} communicate with each other through replicating a chain of commands, which is generally slower than communication with sockets or shared memory. (2) \texttt{Pysyft} is implemented based on \texttt{PyTorch}, and \texttt{PyTorch} usually uses multiply CPU cores to train a model.  According to several conducted experiments, e.g., we observed that when the number of parties is changed in the range of 2-100, \texttt{Pysyft} always uses about ten CPU cores, and the communication between parties is slower than other platforms.}

\section{Discussion}
\label{sec:disc}

\begin{figure*}[htb]
    \centering
    \includegraphics[scale=1]{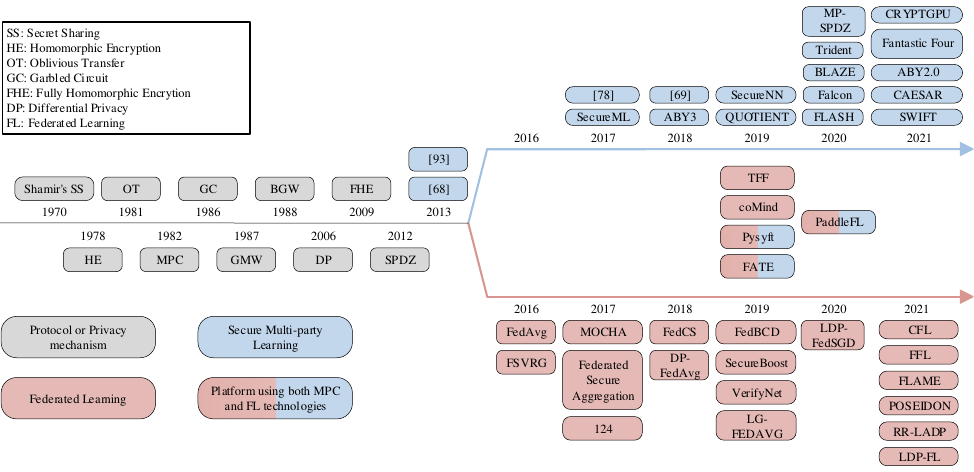} 
    \caption{The development history of the solutions to resolve \TMMPP}
    \label{fig.development}  
\end{figure*}

The solutions to \TMMPP enable multiple data owners to collaboratively \lgp{train} a machine learning model while keeping all the training data private. This would solve the open problem of isolated data sources, which is exacerbated by emerging privacy protection laws and regulations. 
In recent years, this topic has aroused much interest and attention, in both academia and industry.

In this section, we present the development history of the solutions to solve \TMMPP, then discuss the future directions to resolve \TMMPP.


\subsection{Development History}
The development history of the solutions to \TMMPP is summarized in chronological order, as is shown in Figure \ref{fig.development}.
After the concept of MPC was proposed in 1982, the development of MPL was mainly to improve the underlying protocols, although there are some frameworks for training LR models or for secure neural network inference. Until 2017, Mohassel et al. proposed secureML~\cite{mohassel2017secureml}, and MPC-based machine learning frameworks that support the training phase became hotter, especially after 2019. The concept of FL was proposed in 2016 and is still in its infancy. Since 2019, a large number of FL platforms have been developed by many companies to deal with real-world problems or for research. 

\subsection{Future Directions}
In the future, when solving \TMMPP, we should cover a wide spectrum of topics, i.e., improving the security, improving the efficiency, supporting more data distributions, and others.

\subsubsection{\textbf{Improving the security}}
As the assurance frameworks of data, improving the security of the solutions to \TMMPP is still significant in future directions. To improve security, we should countermeasure two key issues: supporting weaker security assumptions, and access control of the global model. 
\begin{itemize} 
    \item \emph{Supporting weaker security assumptions:}
    Currently, major MPL frameworks focus on an honest majority with semi-honest or malicious corruption.
    To protect the security of data in various scenarios, we are supposed to support weaker security assumptions, e.g., the dishonest majority with malicious corruption. 
    The scheme in SPDZ that makes use of MAC upon additive SS to encounter the malicious behaviors in a dishonest majority security model would be a promising technique route to tackle this \hwl{issue}.  
    Furthermore, the centralized server is usually assumed to be honest or semi-honest in \hwl{major} FL frameworks. I.e., the FL frameworks applicable to situations with a malicious server are still seldom studied. A promising solution is combining the solution of secure aggregation \cite{bassily2017practical} and SPDZ, using zero sharing to mask the model updates and using MAC to check the output of the centralized server. 
    
    
    \item \emph{Access control of the global model:} In FL frameworks, the centralized server sends the plaintext global model to all parties or a subset of them in each round. Therefore, each party has an almost equal opportunity to obtain the global model. In the MPL framework, the model can be stored in each party in the form of ciphertext, but any $t$ cooperation can recover the global model, where $t$ is the threshold. In real-world scenarios, however, it is generally required that only some authorized parties, which could pay the money for the data supplied by other parties, can obtain the global model. As a result,
    enforcing access control methods in the MPL or FL frameworks is suitable for the above scenarios; meanwhile, it is also a promising future direction.
\end{itemize}
    


    
\subsubsection{\textbf{Improving the efficiency}} 
Improving the efficiency in the solutions of \TMMPP is one of crucial future directions. 
There are two feasible routes in MPL frameworks: improving online efficiency, and optimizing the training process.

\begin{itemize}
    \item \emph{Improving the online efficiency}: On the one hand, we can reduce the communication computational complexity in online phase through directly optimizing underlying protocols. On the other hand, we can split the expensive cryptographic operations and data independent calculations into offline phase.
    \item \emph{Optimizing the training process}: \lgp{A better initial model will generally reduce the required iterations for getting a high accuracy.} Thus, we can choose a better initial model for training.  The initial model can be obtained by aggregating the local models which \lgp{are pre-trained by each party locally}.
\end{itemize}

\subsubsection{\textbf{Supporting more data distributions}}
The non-IID data has a signicant influence on the training results of FL. 
The multi-task learning~\cite{smith2017federated}\cite{corinzia2019variational} and meta-learning~\cite{li2019differentially}\cite{jiang2019improving}\cite{fallah2020personalized} enable personalized or party-specific modelling. Thereby they are efficient approaches to handling the statistical challenge in FL frameworks. 

\subsubsection{\textbf{Others}}

Other important research directions in the future are as follows:

\begin{itemize}
    \item \emph{Improving dynamic scalability in MPL frameworks}: Most of the optimizations are tailored for a specific number of parties in current MPL frameworks. Especially, the number of parties does not support dynamic changes, which could be led by unexpected crash of any party. Once the number of parties changes during the training, the current MPL frameworks will fail. In order to support the dynamic scalability of the number of parties, Sharmir's SS as an underlying protocol would be a feasible underly protocol to support future MPL frameworks. 
 
    \item \emph{Supporting more complex models}: The current MPL and FL frameworks generally support simpler parametric models, such as LR and LOR, but some non-parametric models, such as decision trees, or more complex parametric models such as LSTM, cannot be supported or are low efficient. 
    In the real-world, these non-parametric and more complex parametric are also widely used.
    Thereby, the solutions of \TMMPP can support more models is also a meaningful future direction.
    \item \emph{Tolerating system heterogeneity}: 
    \lgp{
    Major MPL and FL frameworks are implemented based on the assumption that the communication channels are reliable and each party is equipped with similar computational capability. Once some parties drop out or respond slowly, the frameworks may crash or have significant latency. However, in actual scenarios, the communication channels of some parties could be unreliable, especially in the wireless channels (3G, 4G, 5G, WiFi). In addition, the computational capability of each party may be different due to the variability in hardware (CPU, memory). 
    By referring to the designs in traditional distributed computing technologies, these frameworks may consider to further support the flexible participation or withdrawal of parties and dynamic computing tasks allocation to tolerate the system heterogeneity.}
\end{itemize}

\bibliographystyle{IEEEtran}
\bibliography{survey}

\begin{appendices}

\section{Machine Learning Models}
\label{sec.ml}

In this section, we briefly review the state-of-the-art machine learning models mentioned in Section \ref{sec:frameworks}: LR, LOR, deep neural network (DNN for short), convolutional neural network (CNN for short), support vector machine (SVM for short) and decision tree.

\begin{itemize}

    \item \textbf{LR:}  LR is a fundamental building block of many machine learning algorithms where the relationship among independent and dependent variables is linear. Given a set of data points, it produces a model by fitting a linear curve through the data points. Formally, the linear function can be represented as $g(\mathbf{x}) = \mathbf{x} \cdot \mathbf{w}$, where $\mathbf{x}$ is the input vector and $\mathbf{w}$ is the parameter vector.
    
    The parameters of fitting function are set by continuously calculating the gap between the current fitted value and the actual value, then updating the parameters according to the gradient generated by the gap. LR repeats this iteration several times (until convergence).
    \item \textbf{LOR:} In classification problems with two classes, LOR introduces the logistic function $f(u) = \frac{1}{1+e^{-u}}$ as the activation function to bound the output of the prediction between 0 and 1. Thus the relationship of LOR is expressed as $g(\mathbf{x}) = f(\mathbf{x} \cdot \mathbf{w})$.
    \item \textbf{Support Vector Machine:} SVM is a supervised machine learning model that uses classification algorithms for two-group classification problems. The main idea of SVM is to find a hyperplane that has the maximum margin for all data samples. For the data samples which are linear inseparable, SVM uses kernel methods to project low-dimension samples to high-dimension space to make all samples could be separated by a high-dimension hyperplane.
    \item \textbf{Deep Neural Network:} DNN is one of the supervised learning models, which can learn complex and non-linear relationships among high dimensional data. DNN consists of one input layer, one output layer, and multiple hidden layers, where the output of each layer is the input to the next layer. Each unit in the network is named a neuron. The value of each neuron except for those in the input layer is calculated by a linear function with the input from parameters and the value of neurons from former layers. After the linear function, each neuron is also processed by an activation function (such as ReLu, Sigmod).
    
    \item \textbf{Convolutional Neural Network:} CNN is widely used in image processing and computer vision, which takes pictures represented as matrices as input. The structure of CNN is similar to DNN, but it has multiple additional layers, convolutional layers, and pooling layers. In a CNN, the amount of information of input pictures can get well abridged by the convolutional layers and pooling layers, which utilize the stationary property of images and can greatly reduce the number of parameters while preserving a promising accuracy.

    \item \textbf{Decision Tree:} Decision tree is a non-parametric model, containing nodes and edges. Each interior node corresponds to the partitioning rule; the edges leaving a node correspond to the possible values taken on by that partitioning rule and leaf nodes  correspond to class labels. Given a decision tree, a feature vector is classified by  walking the tree starting from the root node, and using the partitioning rule represented by each interior node to decide which edge to take  until reaches a leaf. The class result will be found in the leaf node.
\end{itemize}

\section{Platforms in Detail}
In this section, we introduce the details of five open-source platforms.
\subsubsection{\texttt{FATE}}

To the best of our knowledge, \texttt{FATE} (Federated AI Technology Enable) is the first open-source industrial-level FL framework presented by WeBank's AI Department in February 2019. It enables multiple companies and institutions to effectively collaborate on training machine learning models in compliance with data security and data protection regulations. 

\texttt{FATE} provides algorithm-level APIs with detailed documents on installation and usage for users to use directly.
It provides a secure computing framework to support various machine learning algorithms, such as LR, LOR, boosting tree \cite{cheng2019secureboost}, NNs, and so on,  which support heterogeneous and homogeneous styles. \texttt{FATE}  also provides various model evaluations, including binary classification,  multi-classification, regression evaluation, and local vs federated comparison. Currently, \texttt{FATE} supports all three FL architectures, including vertical FL, horizontal FL, and federated transfer learning. Additionally, \texttt{FATE} supports both standalone and cluster deployments. 



\subsubsection{\texttt{TFF}}
\texttt{TFF} (TensorFlow Federated), developed by Google, is an open-source framework for federated machine learning and other computation on decentralized data.  

\texttt{TFF} provides a flexible, open framework for locally simulating decentralized computations into the hands of all TensorFlow users. 
It enables developers to simulate the included FL algorithms on their models and data, as well as to experiment with novel algorithms. 
\texttt{TFF} also supports non-learning computation, such as aggregated analytics over decentralized data. 
\texttt{TFF} provides two APIs of different layers: Federated Learning (FL) API and Federated Core (FC) API. FL API offers a set of high-level interfaces, which plug existing Keras or non-Keras machine learning models into the TFF framework. 
With FL API, users can perform FL or evaluation of their existing TensorFlow models, without studying the details of FL algorithms.
FC API comes with a set of lower-level interfaces, is the core of the framework, and also serves as the foundation to built FL.
The interfaces concisely express custom federated algorithms by combining TensorFlow with distributed communication operators within a strongly-typed functional programming environment.

With \texttt{TFF}, developers can declaratively express federated computations, so they could be deployed to diverse runtime environments. 
However, the latest version of \texttt{TFF} currently released only supports horizontal FL without underlying privacy technologies (e.g., HE, MPC, and DP) to protect data security.
Therefore, \texttt{TFF} is only suitable for experimental testing and simulation, and cannot be deployed in a real environment. What's more, \texttt{TFF} only supports single-machine simulation of multiple machines for training models while cannot support cluster deployment.

\subsubsection{\texttt{PaddleFL}}
\texttt{PaddleFL} is an open-source FL framework based on PaddlePaddle~\footnote{\url{https://github.com/PaddlePaddle/Paddle}}, which is a machine learning framework developed by Baidu.
It is mainly designed for deep learning, providing several FL strategies and applications in the fields of computer vision, natural language processing, recommendation, and so on. 

\texttt{PaddleFL} provides a basic programming framework for researchers and encapsulates some public FL datasets that researchers to easily replicate and compare different FL algorithms. 
With the help of the rich model library and pre-trained models of PaddlePaddle, it is also easy to deploy a federated learning system in distributed clusters.
In the design of \texttt{PaddleFL}, it implements secure training and inference tasks based on ABY3~\cite{mohassel2018aby3}, and uses DP as one of the privacy mechanisms. 
As for horizontal FL strategies, \texttt{PaddleFL} implements a variety of different optimization algorithms, such as DP-SGD, FedAvg, and Secure Aggregate, etc. 
For vertical FL strategies, it provides two algorithms, including LOR and NN.  Currently, \texttt{PaddleFL} supports Kubernetes to deploy it easily and open-sources a relatively complete horizontal federated learning version, but vertical federated learning and transfer federated learning is still in the early stages.

\subsubsection{\texttt{PySyft}}
\texttt{PySyft} is an open-source Python library built for privacy-preserving machine learning, which was originally outlined by Pyffel et al.~\cite{ryffel2018generic} and its first implementation was led by OpenMined, one of the leading decentralized AI platforms. 

\texttt{PySyft} is a flexible, easy-to-use library and enables perform private and secure computation on deep learning models. \texttt{PySyft} provides interfaces for developers to implement their algorithms.
It decouples private data from model training, using FL, secure computation techniques (like MPC and HE), and privacy-preserving techniques (like DP) within different deep learning frameworks, such as PyTorch, Keras, and TensorFlow. 
Moreover, \texttt{PySyft} provides a comprehensive step-by-step list of tutorials, designed for complete beginners. These tutorials cover how to perform techniques such as FL, MPC, and DP using \texttt{PySyft}. 
With tutorials, users can learn about all the ways \texttt{PySyft} can be used to bring privacy and decentralization to the deep learning ecosystem.

\subsubsection{\texttt{coMind}}
\texttt{coMind} is an open-source project for jointly training machine learning models with privacy preservation based on TensorFlow. 

It develops a custom optimizer, which implements federated averaging for the Tensorflow to train NNs easily. Besides, it provides a series of tutorials and examples to help users on how to use TensorFlow and config FL.
There are two types of examples provided, including three basic examples and three advanced examples, which introduce how to train and evaluate TensorFlow machine learning models in a local, distributed, and federated way, respectively.
In this project, both the message passing interface and python sockets can implement federated averaging. They take the communication out of TensorFlow and average the weights by a custom hook. 
In addition, Keras framework is used in \texttt{coMind} as the basis of distributed and federated averaging.
However, similar to \texttt{TFF}, \texttt{coMind} does not provide any encryption method.

\subsubsection{Others}
There are other systems \hwl{which try to resolve} \TMMPP. 
JUGO~\footnote{\url{https://jugo.juzix.net/home}}, developed by JUZIX, provides an MPC underlying algorithm platform and integrates an SDK for general MPC algorithms. 
However, JUGO only supports two parties to collaborate in calculating with basic operations, such as addition and comparison.
Hive is a federated learning platform built by Ping An Technology for the financial industry. So far, the platform has not been open-sourced, and the official documents are lacking.
Except for PaddleFL, Baidu develops MesaTEE \footnote{\url{https://anquan.baidu.com/product/mesatee}}, a universal secure computation platform. MesaTEE builds a FaaS (Function-as-a-Service) general computing framework, providing strict and practical privacy and security capabilities.
TF-Encrypted \footnote{\url{https://github.com/tf-encrypted/tf-encrypted}} is a library for privacy-preserving machine learning in TensorFlow. It makes use of the ease of use of the Keras API while training and predicting encrypted data with MPC and HE technologies.
\end{appendices}

\end{document}